\documentclass[aps,pra,onecolumn]{revtex4}
\usepackage{graphicx,amsmath,amssymb,amsthm,bbm}
\usepackage{longtable}
\usepackage{amsfonts}
\usepackage[pdftex]{hyperref}
\usepackage{hypcap} 
\usepackage{amsbsy}
\usepackage{mathrsfs}

\def\refeqn#1{Eq.\ (\ref{Equation::#1})}

\def\reffig#1{Fig. \ref{Figure::#1}}

\def\kpe{\kappa_{E}'}
\def\kpg{\kappa_{G}'}

\def\tx{t_{\chi}}

\def\as{\alpha_S}
\def\das{d \as}

\def\ag{\alpha_G}

\def\al{\alpha_L}

\def\kl{\kappa_L}
\def\kg{\kappa_G}
\def\kc{\kappa_C}
\def\kp{\kappa'}

\def\cmirtHz{\textrm{cm}^{-1}/\sqrt{\textrm{Hz}}}
\def\cmi{\textrm{cm}^{-1}}
\def\RIN{\textrm{RIN}}
\def\tI{\tilde{I}}

\def\tE{\tilde{E}}
\def\bE{\bar{E}}

\def\vt{v_{T}}

\def\erf{\textrm{erf}}

\begin{document}
    \title{Quantum Limits and Robustness of Nonlinear Intracavity Absorption Spectroscopy}
    \author{John K. Stockton}
    \email{john.stockton@gmail.com}
    \author{Ari K. Tuchman}
    \affiliation{Entanglement Technologies, 3723 Haven Ave., Menlo Park, CA}

\date{\today}

\begin{abstract}
We investigate the limits of intracavity absorption spectroscopy with nonlinear media.  Using a common theoretical framework, we compare the detection of a trace gas within an undriven cavity with gain near and above threshold, a driven cavity with gain kept just below threshold, and a cavity driven close to the saturation point of a saturable absorber.  These phase-transition-based metrology methods are typically quantum-limited by spontaneous emission, and we compare them to the empty cavity shotnoise-limited case.  Although the fundamental limits achievable with nonlinear media do not surpass the empty cavity limits, we show that nonlinear methods are more robust against certain technical noise models. This recognition may have applications in spectrometer design for devices operating in non-ideal field environments.
\end{abstract}





\maketitle

\section{Introduction}

Trace gas identification methods play an important role in precision optical metrology with multiple applications having potentially large societal impact.  These include explosives detection \cite{Yinon2007}, manufacturing process control, and medical diagnostics, such as detection of bio-markers in human breath \cite{Breathalyzer2008}.  Many comprehensive reviews have compared various trace detection techniques, considering both practical implementation \cite{Yinon2007} as well as fundamental quantum limits \cite{Ye1998}.  One established technique is absorption detection of a trace gas within an optical cavity.  A laser beam interacts with the absorber multiple times via high-reflectivity cavity mirrors, which effectively enhances the path length and amplifies the minute single pass absorption signal for a fixed volume of gas.

Practical implementations of cavity enhanced detection are aided by the technique of cavity ring-down spectroscopy (CRDS), whereby the absorption is extracted from a fit of the decaying cavity output \cite{Okeefe1988}.  CRDS is largely immune to input laser intensity noise, and its practical utility has led to its prevalence in a variety of commercial applications.  Furthermore, this technique can be made highly parallel by feeding the cavity with multiple frequencies resonant with adjacent cavity modes from a femto-second laser source \cite{Breathalyzer2008}.  Modern mirror and cavity technology achieves finesses of approximately $10^5$ for IR wavelengths with corresponding resolvable absorption coefficients in the range of $10^{-14} \ \cmi$ in a one second timescale \cite{Ye1998}.  This sensitivity achieved under optimum laboratory conditions represents the comparative benchmark for novel approaches.

However, when considering practical field-deployed devices, background technical noise may significantly degrade performance. Therefore, in comparing absorption sensitivity across different  technology platforms, one must consider robustness to non-ideal environments. Measurement schemes with nonlinear media, especially when coupled with modulation, have shown great promise in other fields for providing robustness to such technical fluctuations.  For example, transistors, Josephson bifurcation amplifiers \cite{Devoret2004}, and fluxgate magnetometers \cite{Primdahl1979}, all exhibit robustness in field-deployment. It is thus natural to consider the implications of inserting nonlinear media into intracavity spectroscopy systems.

The simplest addition is an amplifier which can compensate the cavity loss, thereby increasing the effective cavity finesse.   This approach is limited by spontaneous emission as is described via amplifier uncertainty relations in Ref.~\cite{Caves1982}.  This limit can also be cast into the language of the `quantum no-cloning' \cite{Nielsen2000} and the fluctuation-dissipation theorems \cite{Mandel1995}.  The gain element does not need to be driven externally, and a laser cavity (seeded by spontaneous emission) can also be used for sensitive trace-gas absorption spectroscopy.  This was recognized early in the development of laser physics \cite{Hansch1972} and work on broad-gain medium lasers with relatively narrow trace gas absorption lines continues \cite{Baev1999}.  In Ref.~\cite{Kimble1980}, Kimble addresses the specific example of an intracavity, narrow gain-medium laser without external drive, motivated by the observation that a laser has a nonlinear threshold point which is sensitive to intracavity absorption.  In the absence of spontaneous emission, the laser output power curve exhibits a first order phase-transition-like feature at  threshold.  However, as recognized in Ref.~\cite{Kimble1980}, spontaneous emission smooths out the sharpness of the threshold point and limits the sensitivity.  Nevertheless, for absorption measurements in the presence of technical noise, enhanced performance is still possible.  

This idea of enhancing the absorption sensitivity by using the trace gas as a perturbation which kicks the system past a nonlinear bifurcation point (or phase transition) can also be applied to situations without gain.  For example, optical bistability can occur for a saturable absorber within a cavity driven near saturation \cite{Siegman1986, Drummond1980}.  The introduction of additional absorption kicks the system past saturation, thereby amplifying the signal.  While optical bistability is typically discussed as a potential nonlinear transistor element for optical computing, it can also be used for absorption spectroscopy; yet, the performance is once again limited by spontaneous emission.

In this paper we review absorption spectroscopy techniques using nonlinear elements from a unified perspective and derive both fundamental and technical limits.  The potential benefit of such nonlinear techniques depends critically on technology limitations, such as mirror coatings and light sources as well as on technical noise of a non-fundamental nature.  We use a common stochastic formalism to analyze different regions of the allowable phase space with drive and gain as the independent parameters.   In each case, the derived limits are compared to the empty cavity benchmark.  We express the absorption coefficient uncertainty as a function of time, recognizing that measurement noise is not purely white, and dynamical timescales are important.   We conclude by demonstrating that nonlinear techniques can approach, but not improve upon, the fundamental sensitivity limits of empty-cavity methods.  Most importantly, we show that nonlinear approaches can robustly outperform linear counterparts in the presence of technical noise, as might be found when operating outside of a laboratory environment.

Although we will focus on sensitivity under fundamental and technical limits, one important issue we will not treat is selectivity, whereby signal shapes are interpreted in time and frequency space to determine what species or mix of species is present.  There are several ways to imagine incorporating the systems discussed in this paper into a selective apparatus, for example using diffusion metrics with a cavity gradiometer or femtosecond frequency combs with a broadband gain medium.

The structure of this paper is as follows.  We begin by framing the problem and defining the estimation terminology.   Next, cavity terminology is reviewed and the optical shotnoise-limited empty cavity benchmark is presented in these estimation terms.  We then introduce the dynamical modeling which underlies the nonlinear treatment, using a stochastic formalism to represent spontaneous emission. Three different regions of phase space are then treated in turn: gain without drive, gain with drive, and saturable loss with drive.  In the latter case, we focus on nonlinear modulation techniques and compare to CRDS.  For each case, the parameter range for the nonlinear technique with increased robustness in the presence of background noise models is demonstrated.

\section{Estimation Theory}

In order to treat both the linear and nonlinear absorption techniques with a common approach that renders comparisons meaningful in the presence of finite bandwidth noise, we first define our estimation terminology.  An absorption spectrometer produces a spectral plot where, at each wavelength of input probe light, the output is used to determine the absorption coefficient of the sample, denoted $\as$, in units of $\cmi$.   The precision of a single shot measurement as a function of time (at a particular input wavelength) can be represented by the Allan variance \cite{Vanier1989}.  For a scenario with additive white noise only, the Allan variance initially integrates down as $1/T$ where $T$ is the integration time.  The pre-factor on this portion is termed the `short term sensitivity' and is given in units of $\cmirtHz$ and is typically due to white noise sources.  Repeating uncorrelated measurements in time and adding noise in quadrature also results in a noisefloor with the same units.  Due to slower noise sources the Allan variance typically levels off at a level termed the `detection sensitivity' (or bias stability) and is represented in units of $\cmi$.  The sensitivity can be inferred from the short term sensitivity and the stated averaging time.

We estimate the quantity $x$ through the measurement $y$ with the two being related by a function $y=f(x,p_i)$ which is inverted to give $x=g(y,p_i)$.  Here $p_i$ are generic parameters that may be noisy, and thus add to the overall error.  The propagated uncertainty estimation is usually given by adding the uncorrelated noises in quadrature as
\begin{eqnarray}
dx^2 &=& dy^2 \left( \frac{\partial g}{\partial y}\right)^2 + dp_i^2 \left( \frac{\partial g}{\partial p_i}\right)^2.
\end{eqnarray}
With $y$ as the measurement and $x$ as the desired variable, related by $y=f(x,p_i)$, it is typical to refer to the quantity
\begin{eqnarray}
R&=&\frac{\partial f}{\partial x}
\end{eqnarray}
as the responsivity of the detector.  In photodetectors, e.g., this is given in units of Amps per Watt, A/W.  The normalized responsivity is defined as
\begin{eqnarray}
\bar{R}=\frac{R}{f}=\frac{(\partial f/\partial x)}{f}.
\end{eqnarray}
We subsequently do not use $R$ and refer to $\bar{R}$ as the responsivity with the implicit understanding that it is normalized.   The sensitivity of the detector  $dx^2$ can be represented (ignoring $p_i$ terms above) as
\begin{eqnarray}
dx^2 &=& dy^2 \left( \frac{\partial g}{\partial y}\right)^2=\frac{dy^2}{ \left( \frac{\partial y}{\partial x}\right)^2} = \frac{dy^2}{R^2}=\frac{dy^2}{y^2}\frac{1}{\bar{R}^2}
\end{eqnarray}
Since most experiments display a dominant fast noise that must be integrated down until a technical noise limit is met, the typical form of the uncertainty as a function of time (i.e., the Allan variance) is
\begin{eqnarray}
dx^2(t) &=& \frac{1}{\bar{R}^2}\frac{dy^2}{y^2}= \frac{1}{\bar{R}^2}\left(\frac{v_0}{1+\gamma t}+\vt\right)\label{Equation::GenericSolution}
\end{eqnarray}
where $\gamma$ is the fluctuation rate of the noise source (in frequency space the noise being white below $\gamma$ and rolling off above) and $v_0$ is the short time variance of the signal, which may depend on $\gamma$ in certain circumstances.  In the following analysis, $v_0$ can either be due to shotnoise, with the bandwidth determined by the decay of the cavity and $v_0$ inversely related to the intensity, or it can be representative of near-threshold laser dynamics which have larger variances and slower dynamics (but smoothly merges into shotnoise far above threshold).  The variance $\vt$ represents a technical noise source, such as alignment or gain instability in the post-cavity detection. This noise could be represented as $\vt/(1+\gamma_T t)$, but we consider this bandwidth $\gamma_T$ as too slow to be integrated past; hence the denominator is taken as unity.

Generally, non-modulated schemes as presented below take the form of \refeqn{GenericSolution}.  When one ignores $\vt$, there is typically no enhancement in the fundamentally limited sensitivity achieved by adding an extra nonlinear element.  However, the nonlinear element can enhance the responsivity $\bar{R}$ such that when $\vt$ is non-zero, the long term performance improves significantly.  In other words, the nonlinear system is not optimal in a fundamental sense, but makes the system robust to certain types of technical noise.

The above propagation-of-uncertainty style of analysis is typical but not generally optimal.  Instead, we could apply an optimal Bayesian estimator to the problem, utilizing full knowledge of the system parameters \footnote{Note that a full Bayesian approach, i.e., optimized curve fitting, could be applied to CRDS, producing results similar to those of \cite{Stockton2004a}, where the variance in determining the slope, hence absorption, at short times goes as $1/t^3$.}.  In the simple case where the noise and distributions are gaussian and the dynamics linear (or appropriately linearized), the solution for the uncertainty as a function of time (i.e., the Riccati equation solution) typically takes the intuitive form of the estimation variance in \refeqn{GenericSolution}.  Nevertheless, our analysis suffices to evaluate the relative merit between linear and nonlinear absorption techniques.

\section{Single Pass and Empty Cavity Benchmarks}

The simplest absorption measurement, and therefore the most transparent application of the above analysis, uses input power $P_0$ incident on a medium such that the output power $P_1$ is given by
\begin{eqnarray}
P_1&=&\exp((\ag-\al-\as)L)P_0
\end{eqnarray}
where $\ag$ is the gain per unit length, $\al$ is the non-sample loss, $\as$ is the sample loss we are trying to detect, and $L$ is the length of the element.  Thus the  responsivity of the system is given by
\begin{eqnarray}
\bar{R}&\equiv&\frac{\partial P_1/\partial \as}{P_1}=-L
\end{eqnarray}
This is independent of the loss factors, but the sensitivity is not.  The sensitivity is given by
\begin{eqnarray}
\das^2=\frac{1}{\bar{R}^2}\frac{dP_1^2}{P_1^2}=\frac{1}{L^2}\frac{dP_1^2}{P_1^2}
\end{eqnarray}
The fractional power uncertainty from averaging down shotnoise is $dP_1^2(t)=\hbar\omega P_1/t$ \cite{Vanier1989} giving
\begin{eqnarray}
\das^2=\frac{1}{L^2}\frac{\hbar\omega}{t P_1}=\frac{1}{L^2}\frac{\hbar\omega}{t \exp((\ag-\al-\as)L)P_0}
\end{eqnarray}
which, for $\ag=0$ and $\as \ll \al$, is optimized with $L=2/\al$ at
\begin{eqnarray}
\das^2=\frac{\al^2}{4}\frac{\hbar\omega}{t \exp(-2)P_0}\label{Equation::SinglePassSensitivity}
\end{eqnarray}
This tells us to make the medium long enough that we see the effect of the absorption, but not so long that the shotnoise suffers due to the loss of light.  Here we have ignored the effect of spontaneous emission due to any of the elements in the cavity. 

We now extend the single pass treatment to an empty cavity, which represents a more realistic practical implementation.  Advances in mirror coatings and cavity fabrication have reached the state where mirror losses are at the ppm level in the NIR.  In addition to enabling high finesse cavities, these coatings also have relatively high damage thresholds which permit large intracavity intensities for measurement sensitivities approaching the photon shotnoise level.  However, the quoted damage threshold is often significantly above the level where photorefractive processes introduce additional noise \cite{Hall2000}, and 10 kW/cm$^2$ has been shown to be a reasonable upper limit.  Cavity fabrication techniques using materials with low thermal coefficient of expansion such as Zerodur \cite{Tuchman2006} and optimized geometries \cite{Ludlow2007} have also rendered high finesse cavities robust to environmental conditions.   In addition to mirror constraints, intensity limits may also be limited by practical considerations such as power consumption budgets for battery-operated field devices.

We begin by introducing basic optical cavity terminology following closely the notation of \cite{Siegman1986} which is used throughout.  Consider the intensity losses within a Fabry-Perot cavity of length $L$ in one round trip, with each round trip taking time $2L/c$.  The total loss per round trip is given by 
\begin{eqnarray}
\delta&=&\frac{2L\kappa}{c}=\delta_C+\delta_S+\delta_L
\end{eqnarray}
where $\kappa$ gives the total intensity loss rate.  The total empty cavity loss is given as $\delta_C=2L\kappa_C/c=\delta_0+\delta_1+\delta_2$, consisting of the absorptive loss from the mirrors $\delta_0$ and the transmissive loss from one mirror given by $\delta_1=\ln(1/R_1)\approx 1-R_1$ (and a similar term for the other mirror with $\delta_2$), where we have assumed the mirror reflectivity $R_1$ to be close to unity.  We will assume negligible absorption loss, $\delta_0=0$ and a symmetric cavity $\delta_1=\delta_2$, so that $\delta_C=2L\kappa_C/c=2\delta_1$.  The total sample loss is given by $\delta_S=2L\kappa_S/c=2L\as$.  The total additional absorber loss (or gain) is $\delta_L=2L\kappa_L/c=2L\al$, which we allow to be intensity dependent due to saturation. The free spectral range (axial frequency spacing between modes) is $\Delta f_{ax}=c/2L$.  The effective finesse of the cavity is $F_{\textrm{cav}}=2\pi/\delta$, and the linewidth of the cavity (equivalent to the loss rate) is $\Delta f_{\textrm{cav}}= \kappa/2\pi=\Delta f_{ax}/F_{\textrm{cav}}$.

We introduce the dynamical equation for the unitless internal cavity field $E$, where the intensity $I=E^2$ is equal to the number of photons in the cavity mode.  We also add a driving term $\tE_0$ to give \cite{Walls1994}
\begin{eqnarray}
\frac{dE}{dt}&=&-\frac{\kappa(I)}{2}E + \tE_0 \label{Equation::NonSDE}
\end{eqnarray}
The intensity equation is then
\begin{eqnarray}
\frac{dI}{dt}&=&2E \frac{dE}{dt}= -\kappa(I)I + 2\sqrt{I}\tE_0
\end{eqnarray}
In steady state, when all losses are due to the cavity ($\kappa=\kappa_C$) we require the input power ($P_0 \propto \tE_0^2$) to equal the output power ($\hbar\omega\kappa_C I $) which gives $P_0=4\hbar\omega \tE_0^2/\kappa_C$.  When the sample and other losses are included we have the steady state condition, $dI/dt=0$, for the output power
\begin{eqnarray}
P_1&=&\hbar\omega\kappa_C I = \frac{\kappa_C^2}{\kappa(I)^2}P_0=\frac{4 \delta_1^2}{\delta(I)^2}P_0 \label{Equation::CavityIO}
\end{eqnarray}
which can lead to bistability because of the absorption dependence on $I$ through $\delta(I)$.

Ignoring intensity dependence of the absorption, the output power is
\begin{eqnarray}
P_1&=&\frac{4\delta_1^2}{(2\delta_1+2L\as+\delta_L)^2}P_0.
\end{eqnarray}
Thus the responsivity is given by
\begin{eqnarray}
\bar{R}&\equiv&\frac{\partial P_1/\partial \as}{P_1}=\frac{-4L}{\delta}\approx\frac{-2 L}{\delta_1}=-L\frac{2 F}{\pi} \label{Equation::EmptyResponsivity}
\end{eqnarray}
where we assume $\delta_L=0$, and $2\delta_1\gg2L\as$.  The responsivity is enhanced over the single pass result by a factor proportional to the finesse.  Note that if the gain is increased past threshold (so $\delta<0$), a steady state no longer exists.  In practice the saturation of the gain limits the laser output power above threshold and is treated below.

The uncertainty in $\as$ due to shotnoise is given by
\begin{eqnarray}
\das^2&=&\frac{dP_1^2}{(\partial P_1/\partial \as)^2}=\frac{1}{\bar{R}^2}\frac{dP_1^2}{P_1^2}=\frac{\delta^2}{16 L^2}\frac{\hbar\omega}{t P_1}\nonumber\\
&=&\frac{(2\delta_1+\delta_{L})^4}{16 L^2}\frac{\hbar\omega}{t 4\delta_1^2 P_0}
\end{eqnarray}
where we have assumed  $\as$ is small.  This is minimized for $\delta_1=\delta_{L}/2=\al L$ at
\begin{eqnarray}
\das^2&=&\frac{\hbar\omega \delta_{L}^2}{L^2 P_0 t}=\frac{\hbar\omega 4\al^2}{P_0 t}
\end{eqnarray}
Again the optimal output power is not much smaller than the input, with $P_1=P_0/4$.  Note that the sensitivity is only a factor of order unity from the optimized single pass case \refeqn{SinglePassSensitivity}.  The difference however is that the cavity measurement is much more compact: we can have the same sensitivity for a total amount of trace gas that is $1/F_{\textrm{cav}}$ times smaller, in a container that is $1/F_{\textrm{cav}}$ times smaller.

We next consider the propagation of input `relative intensity noise' defined as $\RIN = dP_0^2/P_0^2$,
\begin{eqnarray}
\das^2&=&\frac{dP_0^2}{(\partial P_0/\partial \as)^2}=\frac{dP_0^2}{(4 LP_0/\delta)^2}\approx\frac{\delta_1^2}{4 L^2}\RIN.
\end{eqnarray}
We see that the cavity provides robustness to input intensity noise by a factor of the finesse squared in the variance.  Note that input fluctuations at frequencies larger than the cavity linewidth will be low-pass filtered by the cavity as well.

We also note that if there are technical loss (or gain) fluctuations that appear symmetrically to the absorber, then they cannot be distinguished, hence the error would propagate trivially as $d\as^2=d\al^2$ for loss.  Thus, to benefit from inserting a medium into the cavity, the loss stability needs to be better than the desired estimation sensitivity.

In quadrature sum, the form of the uncertainty as a function of time (ignoring non-mirror loss, gain/loss fluctuations, and making $\delta_1$ as small as possible) can be written as
\begin{eqnarray}
\das^2&=&\frac{\kappa_C^2}{4 c^2}\left(\frac{1/I}{1+\kappa_C t}+\RIN\right)  \label{Equation::EmptyUncertainty}
\end{eqnarray}
where $I$ is the internal photon number and we have used $P_1=\hbar\omega\kappa_C I$ and $dP_1^2=\hbar \omega \kappa_C P_1/(1+\kappa_C t)$.  This form for the shotnoise results from filtering the detector output at the bandwidth of the cavity, as fluctuations of the absorber cannot be observed at higher frequencies.   Note that for CRDS, the $\RIN$ component due to input laser intensity fluctuations largely disappears and, in the absence of any other noise sources, we are left with the shotnoise limit.

\section{Stochastic Quantum Dynamics for Nonlinear Methods}

So far we have only included optical shotnoise as a fundamental noise source.  We have ignored spontaneous emission from and saturation of the trace gas, which is appropriate for small amounts of trace gas having broad molecular transitions with small cross sections and large saturation intensities.  When additional scattering media is included however, we must include both saturation and spontaneous emission. To treat these noise sources we build on the simplified treatment of \refeqn{NonSDE} and use a stochastic differential equation (SDE) for the field, with the random Wiener increment representing the spontaneous emission (see \cite{GardinerBook1985, Carmichael2003} for comprehensive physical treatments of stochastic methods). This treatment for gain within a cavity follows many previous treatments \cite{Kimble1980, Drummond1980, Mandel1995}, however much of this section builds upon the analysis of Ref.~\cite{Kimble1980} in particular.

The SDE with saturable gain is given by
\begin{eqnarray}
dE_i(t)&=&\left(\frac{1}{2}\left(\frac{\kg}{1+|E|^2/I_s}-\kp\right)E_i+\tE_{0,i}\right)dt+\sqrt{2Q(I)}dW_i(t)\label{Equation::GainSDE}
\end{eqnarray}
where the intensity (photon number) is $I=|E|^2=E_1^2+E_2^2$, $\kg=\ag c$, $I_s$ is the saturation intensity of the gain medium, $\kp=\as c +\delta_1 c/L$, $\tE_{0,i}$ is related to the input power as before, and the Wiener increment $dW_i$ is a Gaussian random variable of mean zero and satisfying $\langle dW_i(t) dW_i(t') \rangle = dt \delta(t-t')$.  This SDE is often recognized as a form of the nonlinear van der Pol equation.  The SDE's for the fields can be recast into SDE's for the intensity and phase using the change of variables
\begin{eqnarray}
E_1&=&\sqrt{I} \cos(\phi)\\
E_2&=&\sqrt{I} \sin(\phi)\\
dI d\phi &=& 2 dE_1 dE_2,
\end{eqnarray}
where the latter comes from the appropriate Jacobian, and the stochastic change of variables rule \cite{GardinerBook1985} to give
\begin{eqnarray}
d I (t)&=&\left(\left(\frac{\kg}{1+I/I_s}-\kp\right) I +2\sqrt{I} (\tE_{0,1}\cos(\phi)+\tE_{0,2}\sin(\phi))+4Q(I) \right)dt\nonumber\\
&&+\sqrt{8Q(I)I}dW_I(t)\\
d\phi(t)&=&-\frac{1}{\sqrt{I}}\left( \tE_{0,1}\sin(\phi)-\tE_{0,2}\cos(\phi) \right)dt+\sqrt{\frac{2Q(I)}{I}}dW_{\phi}
\end{eqnarray}
Although generally the spontaneous emission factor $Q$ is a function of $I$, it becomes independent of $I$ if there is a dominant external pumping process, as in a laser.  Then the form of $Q$ is related to the gain via $Q=2\kg$, which can be calculated by equating the laser cavity photon number with the atomic number distribution in thermal equilibrium \cite{Carmichael2003, Kimble1980}.

These Langevin equations are cast into the form of a Fokker-Planck equation describing the evolution of the probability distribution as follows \cite{Mandel1995} 
\begin{eqnarray}
\frac{\partial p(E_1, E_2, t)}{\partial t }&=& \sum_{i=1}^{2} \partial \left(-A_i(E_i)p(E_1, E_2, t)+\frac{\partial(Q(I) p(E_1, E_2, t))}{\partial{E_i}}\right)/\partial{E_i}
\end{eqnarray}
where $A_i=\frac{1}{2}\left(\frac{\kg}{1+|E|^2/I_s}-\kp\right)E_i+\tE_{0,i}$ and the two noise sources are assumed to be independent.  The steady state equation is
\begin{eqnarray}
\frac{\partial p(E_1, E_2, t)}{\partial t }&=& \sum_{i=1}^{2} \frac{\partial J_i}{\partial E_i}=0\\
J_i&=&\left(-A_i+\frac{\partial Q(I)}{\partial E_i}\right)p(E_1, E_2, t)+Q(I) \frac{\partial p(E_1, E_2, t)}{\partial E_i}
\end{eqnarray}
where $J_i$ represent the probability currents of the system which must conserve total probability.  If $Q$ is independent of $I$, we solve this by setting the probability currents equal to zero, $J_i=0$ and integrating by separation of variables \cite{Mandel1995}.  In the more general case where $Q$ is a function of $I$, however, $\partial (A_1/Q(I))/\partial E_2 \neq \partial (A_2/Q(I))/\partial E_1$, and this solution technique is inapplicable.  (When considering optical bistability, we will let $Q(I)=Q_0 I/(I+I_s)$.)

Assuming constant $Q$, $\tE_{0,2}=0$, and $\tE_{0,1}=\tE_0$, we have
\begin{eqnarray}
p_s(E_1, E_2)&=&\exp\left[\left(\tE_0 E_1 - \frac{\kp}{4}(E_1^2+E_2^2)+\frac{\kg I_s}{4}\log(1+(E_1^2+E_2^2)/I_s)\right)/Q\right]\\
p_s(I, \phi)&=&\exp\left[\left(\tE_0\sqrt{I}\cos(\phi) - \frac{\kp}{4}I+\frac{\kg I_s}{4}\log(1+I/I_s)\right)/Q\right]\\
p_s(I)&=&\int_0^{2\pi} p_s(I,\phi) d\phi \nonumber\\
&=& \exp\left[\left(-\kp I+\kg I_s\log(1+I/I_s)\right)/4Q\right] I_{MB0}(\tE_0 \sqrt{I}/Q)\label{Equation::Distribution}
\end{eqnarray}
where we have used the definition of the zero-order modified Bessel function
\begin{eqnarray}
I_{MB0}(x)&=&\frac{2}{\pi}\int_0^{2\pi}\exp[x\cos(\phi)]d\phi\nonumber\\
&\approx&\frac{2}{\pi}\int_0^{2\pi}\exp[x(1-\phi^2/2)]d\phi\nonumber\\
&=&\sqrt{\frac{8}{\pi}}\frac{\exp(x)}{\sqrt{x}}
\end{eqnarray}
and $x\gg 1$ in the approximation.

Now we calculate the moments
\begin{eqnarray}
\langle I^n \rangle &=& \frac{\int_0^{\infty} I^n p_s(I,\phi) dI}{ \int_0^{\infty} p_s(I,\phi) dI}
\end{eqnarray}
and their derivatives with respect to $\kp$ which are
\begin{eqnarray}
\frac{d\langle I^n \rangle}{d\kp} &=& \frac{\int_0^{\infty} (-I/4Q) I^n p_s(I) dI}{\int_0^{\infty} p_s(I) dI}+ \frac{\int_0^{\infty} I^n p_s(I) dI\int_0^{\infty} (I/4Q) p_s(I) dI}{(\int_0^{\infty} p_s(I) dI)^2}\nonumber\\
&=& -(\langle I^{n+1} \rangle-\langle I^{n} \rangle\langle I \rangle)/4Q
\end{eqnarray}
Thus, the responsivity is given by
\begin{eqnarray}
\bar{R}&=&\frac{d\langle I \rangle/d\kp}{ \langle I \rangle}\frac{d\kp}{d\as}=\frac{-c}{4Q} \frac{\langle I^{2} \rangle-\langle I \rangle^2}{\langle I \rangle}.\label{Equation::ResponsivityEquation}
\end{eqnarray}
Note that when the drive is very large, and the system is above saturation, the responsivity returns to that of the empty cavity.

At this point we numerically analyze the entire phase space with the drive and gain, $\tE_0$ and $\kg$, as the free parameters.  For the near threshold case, we simulate the responsivity from \refeqn{ResponsivityEquation} using the distribution in \refeqn{Distribution}.  The result is shown in \reffig{Heart}, where the responsivity peaks just above threshold.  The threshold curve in \reffig{Heart}C is shown to have a nonlinear kink at this point.  But because of non-zero spontaneous emission, the curve is smoothed and the responsivity becomes non-infinite, which is the primary result of \cite{Kimble1980}.  The drive term further smooths out the curves, reducing the responsivity accordingly, as is shown below.  (See Ref.~\cite{Li1994} for an exploration of the dependence of the threshold point on the drive field.)

\begin{figure}[t]
\capstart
\includegraphics[width=3.0in]{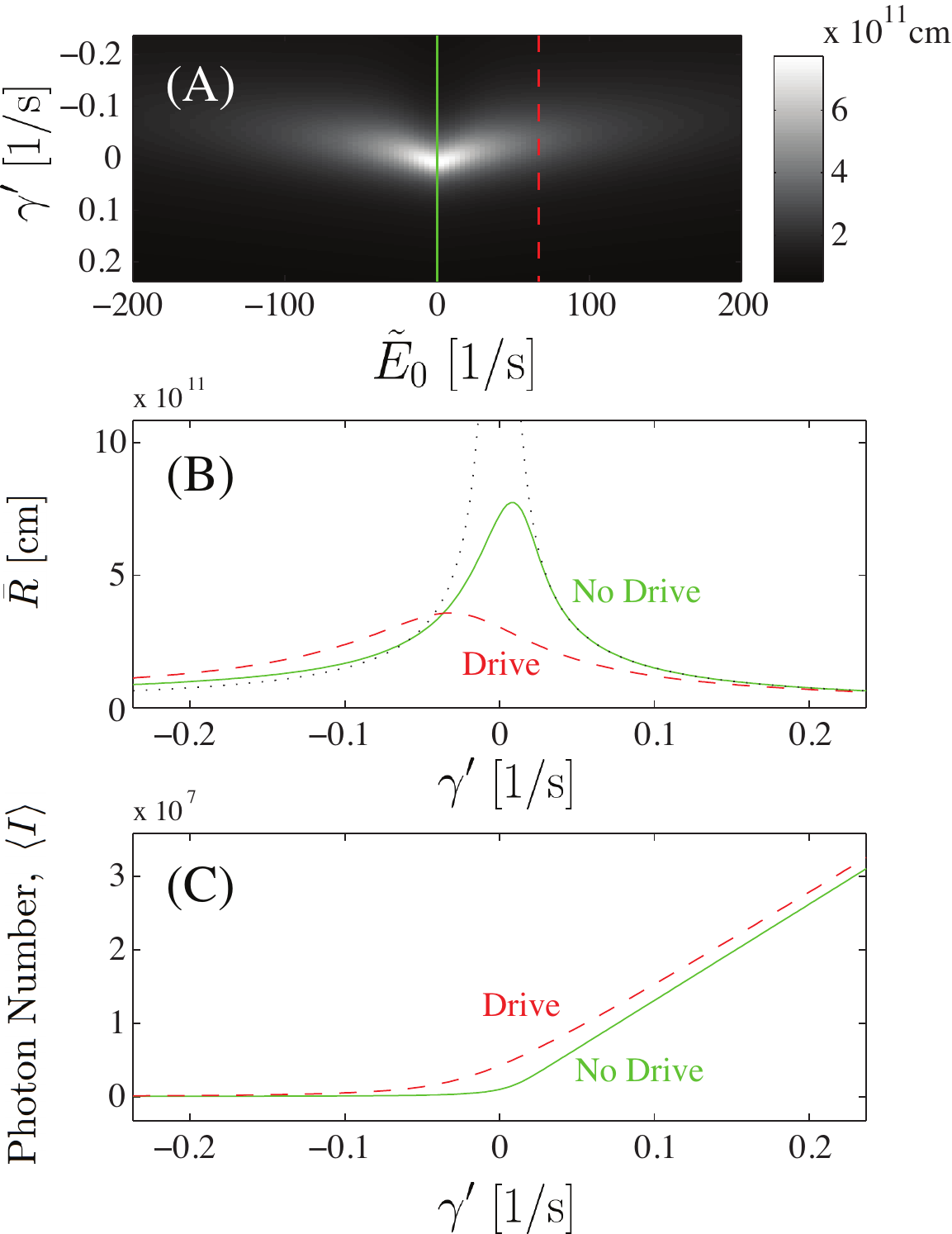}
\caption{(A) Normalized responsivity $\bar{R}$ as a function of drive field and gain.  (B) $\bar{R}$ versus gain for zero (non-zero) drive field.  Green solid (red dashed) line refers to the cross-sections plotted in (A). The dotted black lines diverging near saturation give the zero-drive responsivity without the spontaneous emission limit ($c/2\gamma'$). (C) Intensity as a function of gain for zero drive field (green, solid line, sharp kink), and non-zero drive field (red, dashed line, smoothed kink).  Parameters are $\eta_0=10^6$, $\delta_1=10^{-5}$, $L=1$ m.} \label{Figure::Heart}
\end{figure}

We now consider three different regions of the phase space parameterized by gain and drive: i) laser above threshold without drive, ii) laser below threshold with drive, and iii) absorptive bistability with drive.  We compare each of these schemes to the empty cavity benchmark while considering both quantum and technical limits.

\subsection{Gain without drive}

Previous demonstrations of intracavity gain medium for absorption spectroscopy have utilized a broadband gain medium placed inside the cavity mode.  These have included dye lasers, Ti-Sapphire, fiber lasers and diode lasers, with the dye lasers demonstrating the best sensitivity \cite{Baev1999}.  For compact devices Helium Neon lasers are attractive, specifically in leveraging existing hardware from a commercial ring laser gyroscope, where the gain medium is integrated into a high finesse cavity.  As will be discussed below, increased sensitivity is obtained for large laser threshold photon number and therefore a tapered amplifier gain medium with saturation intensities at the 1 W level can be considered.  Using semiconductor lasers may also facilitate gain stabilization because of the phase dependence on carrier number \cite{Yariv1967}.  However, careful mode matching of the beam to the micron sized active region of the laser without introducing excessive additional loss is required \cite{Sortais2007}.  Nevertheless, advances in high power compact lasers have rendered robust intracavity laser absorption measurements encouraging.

In this section, we keep the gain just sub-threshold such that $\kg\approx\kp$, but $\kg<\kp$.  The gain is also kept low enough that the overall intensity is below saturation of the gain element.  This section essentially reproduces the results of \cite{Kimble1980}, but with an emphasis on surpassing the empty cavity performance by a given factor in a given amount of time in the presence of additional technical noise.

We consider the limit of $I\ll I_s$, where the SDE of \refeqn{GainSDE} becomes
\begin{eqnarray}
dE_i&=&\left(\frac{1}{2}\left(\frac{\kg}{1+|E|^2/I_s}-\kp\right)E_i+\tE_{0,i}\right)dt+\sqrt{2Q}dW(t)\nonumber\\
&\approx&\left(\frac{1}{2}(\kg(1-|E|^2/I_s)-\kp)E_i+\tE_{0,i}\right)dt+\sqrt{2Q}dW(t)\nonumber\\
&=&\left((\kg/2-\kp/2-|E|^2(\kg/2 I_s))E_i+\tE_{0,i}\right)dt+\sqrt{2Q}dW(t)\nonumber\\
&=&\left((\gamma'-\beta'|E|^2)E_i+\tE_{0,i}\right)dt+\sqrt{2q'}dW(t)
\end{eqnarray}
and we have translated into the notation of \cite{Kimble1980} using
\begin{eqnarray}
\gamma'&=&(\kg-\kp)/2\\
\beta'&=&\frac{\kg}{2 I_s}\\
q'&=&Q
\end{eqnarray}
We write the above steady state distribution
\begin{eqnarray}
p_s(I)&=&\exp\left[\left(\gamma'I/2 - \beta' I^2/4\right)/q'\right] I_{MB0}(\tE_0 \sqrt{I}/q')\\
p_s(\tI)&=&\exp\left[-(\tI-a)^2/4\right] I_{MB0}(\tE_0 \sqrt{\tI/\sqrt{q'^3\beta'}})
\end{eqnarray}
where $\tI=I\sqrt{\beta'/q'}$ and $a=\gamma'/\sqrt{\beta'q'}$.  Without the drive field, the intensity moments are
\begin{eqnarray}
\langle \tI \rangle &=& a + \frac{2\exp(-a^2/4)}{\sqrt{\pi}\left(1+\erf(a/2)\right)}\\
\langle \Delta \tI^2\rangle &=& 2 - \frac{2a\exp(-a^2/4)}{\sqrt{\pi}\left(1+\erf(a/2)\right)}- \frac{4\exp(-a^2/4)}{\pi \left(1+\erf(a/2)\right)^2}
\end{eqnarray}
At threshold ($a=0$), $\langle \tI \rangle = \frac{2}{\sqrt{\pi}}\approx 1.128$ and $\langle \Delta \tI^2\rangle = 2 - \frac{4}{\pi}\approx 0.727$.  The intracavity photon number (intensity) at threshold is defined as
\begin{eqnarray}
\eta_0&=&\sqrt{\frac{q'}{\beta'}} \frac{2}{\sqrt{\pi}}=\sqrt{\frac{2 I_s Q}{\kg}} \frac{2}{\sqrt{\pi}}=\frac{4}{\sqrt{\pi}}\sqrt{I_s}
\end{eqnarray}
for a fully inverted medium with $Q=2 \kg$.  The below saturation approximation $\langle I \rangle \ll I_s$, implies $a\ll\eta_0$ and $\langle I \rangle\ll \langle \Delta I^2 \rangle$.  Furthermore, if we assume the below saturation limit, the above threshold limit $a \gg 1$, and $\eta_0\gg 1$, we have
\begin{eqnarray}
\kp&\approx&\kg\\
\gamma'&\approx& \kp\frac{4 a}{\eta_0\sqrt{\pi}}=\kp \frac{a}{\sqrt{I_s}}.
\end{eqnarray}
Thus the following limits are equivalent
\begin{eqnarray}
1\ll& a &\ll \eta_0\\
\frac{\kp}{\sqrt{I_s}}\ll& \gamma' &\ll \kp\\
\sqrt{I_s}\ll& \langle I \rangle &\ll I_s
\end{eqnarray}

In the limit of large $a \gg 1$, the normalized quantities are
$\langle \tI \rangle = a$ and $\langle \Delta \tI^2 \rangle = 2$, which give
\begin{eqnarray}
\langle I \rangle &=&a \eta_0 \frac{\sqrt{\pi}}{2}=2 I_s \gamma'/\kp\\
\langle \Delta I^2 \rangle &=& 2 \eta_0^2 \frac{\pi}{4} = 8 I_s.
\end{eqnarray}
The responsivity calculated from these moments is essentially the same as in \refeqn{EmptyResponsivity}
\begin{eqnarray}
\bar{R}&=&\frac{d\langle I \rangle/d\as}{\langle I \rangle}=\frac{-c}{2\gamma'}\nonumber\\
&=&\frac{-2L}{\delta}=-\frac{2L}{\delta_1}\left(\frac{\delta_1}{\delta}\right)\label{Equation::GainResp}
\end{eqnarray}
remembering that $\gamma'$ depends on $\kp$ which depends on $\as$. The enhancement factor over the empty cavity, \refeqn{EmptyResponsivity}, is $\delta_1/\delta$ which can be made infinite at $a=0$.  However, this result is invalid unless $a \gg 1$.  When the full equation is used, the responsivity, as displayed in \reffig{Heart}B, peaks at
\begin{eqnarray}
\bar{R}&=&\frac{d\langle I \rangle/d\as}{ \langle I \rangle}=-\frac{2L}{\delta_1}  \left( \frac{0.346}{4} \sqrt{I_s}\right)
\end{eqnarray}
for $a=1.28$ (the normalized function $\frac{d\langle \tI \rangle/d a}{ \langle \tI \rangle}$ is maximized at $0.346$ for $a=1.28$) \cite{Kimble1980}. Thus, the responsivity enhancement factor over an equivalent cavity without a gain element is nearly $\sqrt{I_s}$.

We now determine the sensitivity $\das^2$.  We perform a linearization of the dynamics which allows us to easily apply standard estimation theory techniques.   We start with the full nonlinear problem with $Q$ constant as given in \refeqn{GainSDE}.  For the linearization, we assume $E_i=\bE_i+b_i$, with  $|b_i|\ll\bE_i$, where $\bE_i$ solves the equation without noise,
\begin{eqnarray}
0&=&\frac{1}{2}\left(\frac{\kg}{1+(\bE_1^2+\bE_2^2)/I_s}-\kp\right)E_i+\tE_{0,i}.
\end{eqnarray}
We also ignore all $b_i^2$ terms and assume $\tE_{0,2}=0$, $\bE_2=0$ and consider only $b_1$.  These approximations give
\begin{eqnarray}
dE_1&=&d(\bE_1+b_1)=d b_1\nonumber\\
&=&-\frac{1}{2}\left(\frac{\kg}{(1+\bE_1^2/I_s)}\left(\frac{2\bE_1^2/I_s}{(1+\bE_1^2/I_s)} -1\right)+\kp\right)b_1 dt+\sqrt{2Q}dW_i(t)
\end{eqnarray}
In the far below saturation limit $\bE_1^2/I_s\ll1$ and the dynamics are
\begin{eqnarray}
d b_1(t)&=&\frac{1}{2}\left(\kg(1-3\bE_1^2/I_s)-\kp\right)b_1dt+\sqrt{2Q}dW_i(t)
\end{eqnarray}
With no driving field, $\bE_1^2\approx I_s(\kg-\kp)/\kg$ (need above threshold $\kg>\kp$) and
\begin{eqnarray}
d b_1(t)&=&-\left(\kg-\kp\right)b_1dt+\sqrt{2Q}dW_i(t)\nonumber\\
&=&-2\gamma'b_1dt+\sqrt{2Q}dW_i(t)
\end{eqnarray}
which is an Ornstein-Uhlenbeck process with mean zero and variance $\langle \Delta b_1^2 \rangle  = 2Q/4\gamma'$.  Because we have assumed that the system is above threshold, $\kg>\kp$, this equation is always damped.  Whereas in the full equation the $\gamma'$ term acts as a gain, in the linearized form it turns into a damping.  Because $I=\bE_1^2+2\bE_1 b_1+b_1^2\approx \bE_1^2+2\bE_1 b_1$ with $\bE_1$ a constant, this gives
\begin{eqnarray}
\langle \Delta I ^2 \rangle_0 &=& 4\bE_1^2 \langle \Delta b_1^2 \rangle = 4 \frac{\gamma'}{\beta'}\frac{2Q}{4\gamma'}=\frac{2 Q}{\beta'}=8 I_s
\end{eqnarray}
using $\kg=\kp$.  To check the validity of the assumption we also need
\begin{eqnarray}
1&\ll& \frac{\bE_1^2}{\langle \Delta b_1^2 \rangle}=2a^2,
\end{eqnarray}
thus, the linearization is valid in the regime $1\ll a \ll \eta_0$.

The relative intensity uncertainty as a function of time is
\begin{eqnarray}
\frac{\langle \Delta I ^2 \rangle (t)}{\langle I \rangle^2} &=& \frac{\langle \Delta I ^2 \rangle_0/\langle I \rangle^2}{1+2\gamma' t/2} =\frac{8 I_s/(2 I_s \gamma'/\kp)^2}{1+\gamma' t} =\frac{2\kp^2/ I_s \gamma'^2}{1+\gamma' t}\label{Equation::GainNoise}
\end{eqnarray}
where $\gamma'$ is the effective bandwidth of the fluctuations.  Comparing this to the shotnoise limit of a coherent state within the cavity, which for output $P_1=\hbar\omega\kp \langle I \rangle$ gives
\begin{eqnarray}
\frac{\langle \Delta I ^2 \rangle (t)}{\langle I \rangle^2} &=& \frac{\kp\hbar\omega P_1/P_1^2}{1+\kp t}=\frac{1/ \langle I \rangle}{1+\kp t}=\frac{\kp/ 2I_s \gamma'}{1+\kp t}
\end{eqnarray}
where $\kp$ is the cavity linewidth.  Below saturation $\gamma' \ll \kp$, the shotnoise variance is less than the fluctuation variance above and negligible.  This shows how the laser case limits to the empty cavity case as $\gamma'$ approaches $\kp$.

We now systematically compare the performance of the total gain estimation uncertainty (from \refeqn{GainResp} and \refeqn{GainNoise} with technical noise added in quadrature) to the empty cavity uncertainty of \refeqn{EmptyUncertainty}, which are given by
\begin{eqnarray}
(\das^2)_{G}&=&\frac{4\gamma'^2}{c^2}\left(\frac{2\kpg^2}{I_s \gamma'^2}\frac{1}{1+\gamma' t}+\vt\right)\label{Equation::GainUncertainty2}\\
(\das^2)_{E}&=&\frac{\kpe^2}{4c^2}\left(\frac{1}{I_E}\frac{1}{1+\kpe t}+\vt\right)\label{Equation::EmptyUncertainty2}
\end{eqnarray}
where the subscripts G and E denote gain and empty cavity respectively.  In the empty cavity case, we have replaced the $\RIN$ with $\vt$ for the sake of generality.  

When comparing the two cases we make the following assumptions.  First, we assume cavity dimensions are the same but set $\kpg\gg \kpe$ by an order of magnitude to account for extra loss in the cavity with gain due to mode-matching through the gain element.  We assume that there are no technical gain or loss fluctuations that would affect the nonlinear case more severely, and we also assume that $\vt$ (due to an unspecified technical noise source) is the same for both.   We optimize the internal intensity of the gain case between the ranges of validity of the expressions used $\sqrt{I_s}\ll \langle I \rangle \ll I_s$, but also limit it by the same mirror threshold.  The mirror threshold intensity $I_M$ in units of photon number (and $I_{M,u}$ in regular intensity units) is $I_M=I_{M,u} \frac{A L}{\hbar \omega c}$ and we restrict the gain case by $\langle I \rangle <I_M$.  We also set the empty cavity intensity at this limit, $I_E=I_M$.  

For short times, fluctuations in the gain case are worse than shotnoise, but with $\vt=0$ and an equivalent number of internal photons, the long time (averaging down as $1/t$) limit is approximately the same for the two cases.

\begin{figure}[t]
\capstart
\includegraphics[width=3.0in]{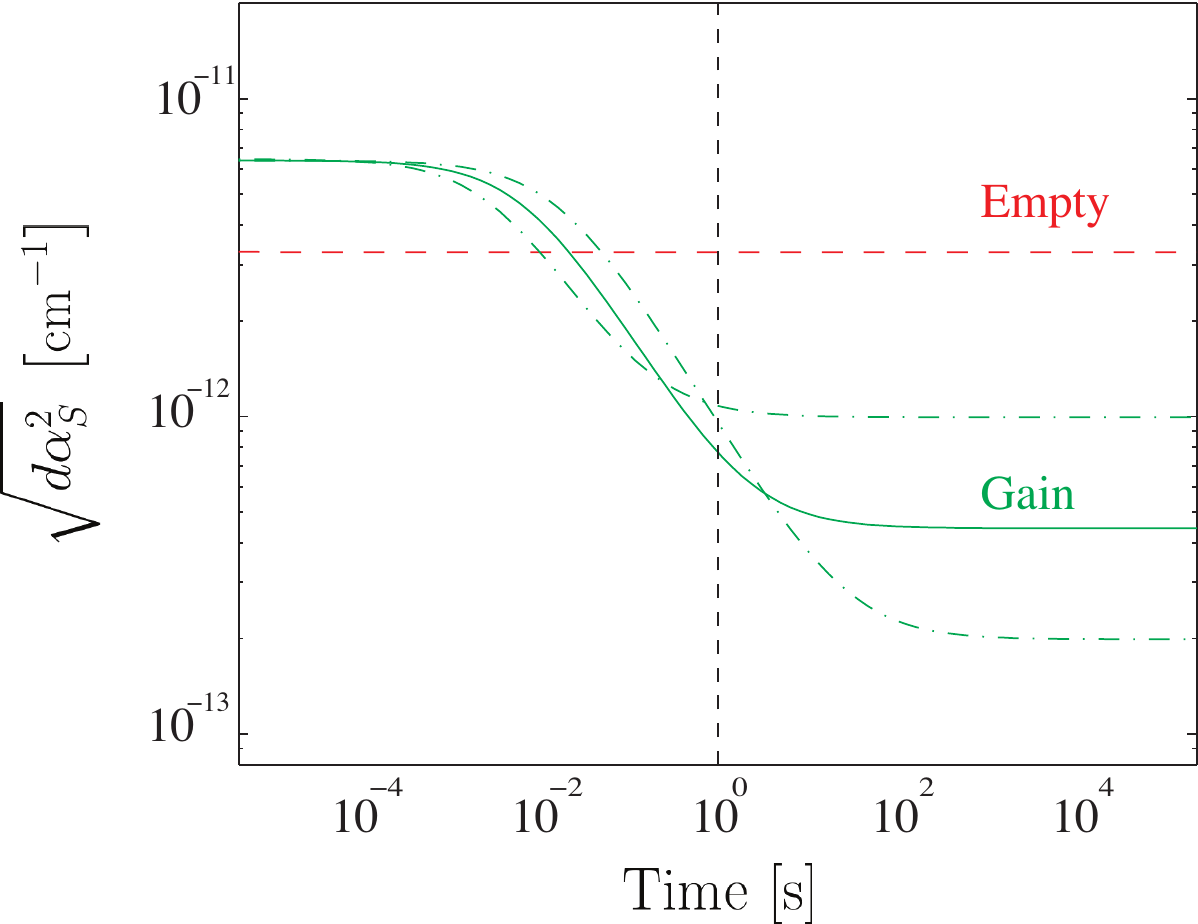}
\caption{Estimation performance for a cavity with gain (green solid line) and without gain (red dashed line).  The gain case eventually integrates down past the no-gain case in the presence of technical fluctuations due to the enhanced responsivity.  The gain case is optimized over $a$ for peak performance at one second (vertical dashed line), and two curves with $a$ above and below the optimum value are also shown (green dashed-dotted line). Parameters are $\eta_0=10^6$, $\delta_{1,E}=10^{-5}$,  $\delta_{1,G}=10^{-4}$, $\lambda=1.064 \ \mu$m, $A=\pi (1 \ \textrm{mm})^2$, $L=1$ m, $I_{M,u} = 10$ kW/cm$^2$, $\vt=10^{-9}$.}\label{Figure::Time}
\end{figure}

\reffig{Time} shows the estimation performance as a function of time.  For the gain case, the uncertainty starts at the variance level of the cavity dynamics, integrates down at times greater than the inverse of the fluctuation rate until it hits the technical noise floor set by $\vt$.  (The shape of the empty cavity case is the same, but relatively flattened because of the high value of the internal intensity, as $\vt \gg 1/I_E$.)  The maximum improvement occurs at long times 
\begin{eqnarray}
\frac{(\das^2)_{G}}{(\das^2)_{E}}&=&\frac{16 \gamma'^2}{\kpe^2}.
\end{eqnarray}
We find that one should make $\gamma'$ small but this slows the fluctuation rate and reaching this limit takes too long.  Instead, we consider the case where we pick a specific time (one second) and optimize \refeqn{GainUncertainty2} at that time over the variable $\gamma'$.  If the optimal $\gamma'$ falls outside the allowed range $\kpg/\sqrt{I_s}\ll\gamma'\ll\kpg$ (or the intensity goes greater than mirror threshold), we fix it at the closest bound.

Using this approach, we plot $(\das^2)_{G}(t)$ in \reffig{Vt0}A which is optimized at $\gamma'$ corresponding to the internal intensity shown in \reffig{Vt0}B.  For sufficiently large $\vt$, the gain case can significantly outperform the empty cavity limit of  $(\das^2)_{E}(t)$, although both clearly get worse in an absolute sense as $\vt$ increases.  In this sense, the gain technique is more robust to the technical noise than the empty cavity approach.

\begin{figure}[t]
\capstart
\includegraphics[width=3.0in]{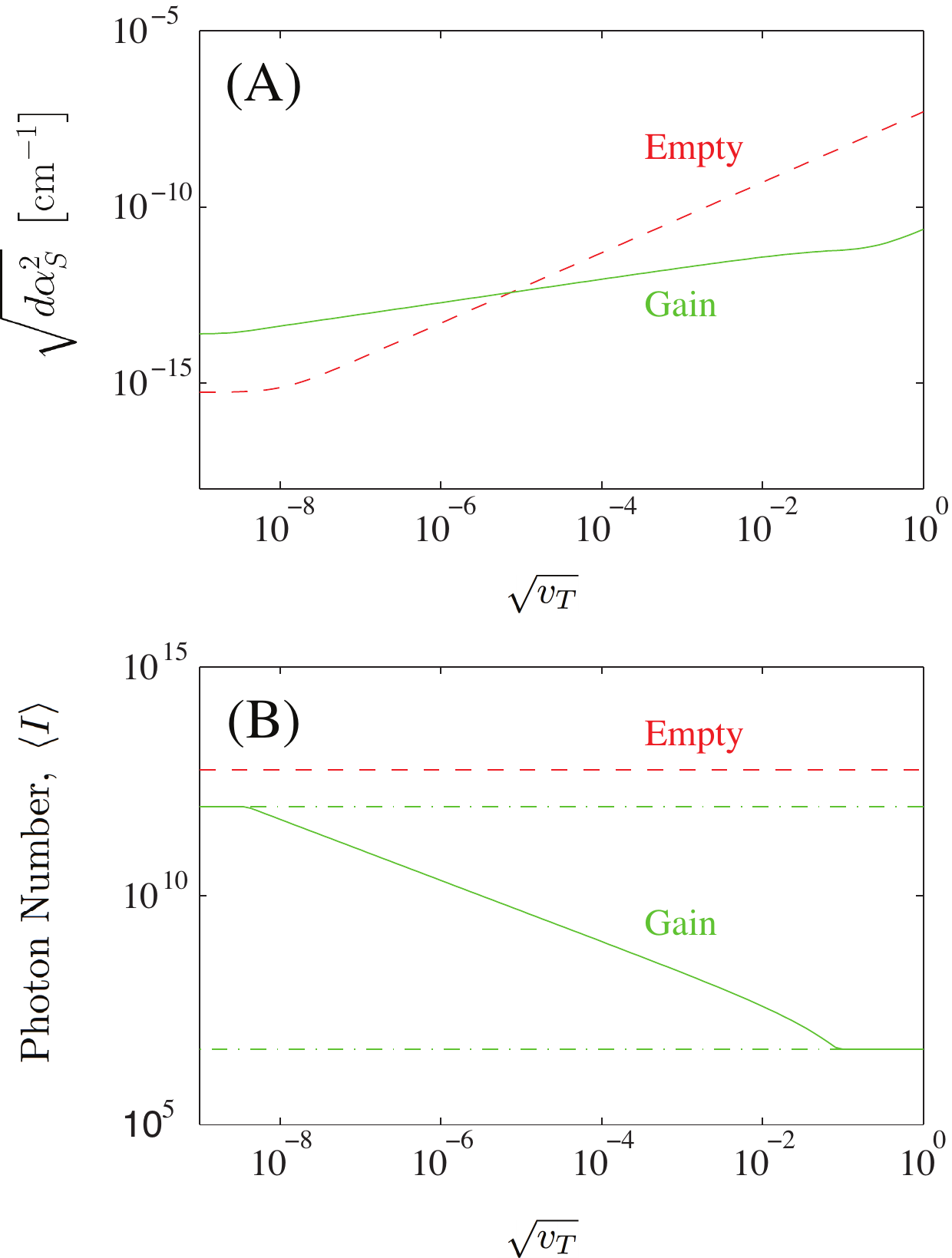}
\caption{(A) Estimation performance in one second for the empty cavity at the mirror limit intensity (red dashed line), and the cavity with gain at optimized $a$ (green solid line) .  (B) The green solid line gives the internal photon number corresponding to the optimal $a$, with $a$ bounded by $5\ll a \ll \eta_0$ ($\sqrt{I_s} \ll \langle I \rangle \ll I_s$, green dashed-dotted lines).  The red dashed line gives the internal photon number used for the empty cavity case (at the mirror limit, $I_{M,u}$). Parameters are the same as in \reffig{Time}. } \label{Figure::Vt0}
\end{figure}

We analyze the robustness improvement factor as follows. At some time $\tx$ we desire the sensitivity improvement with gain to be some factor $\chi$ better than the no gain performance.  
We define the time that  the gain-sensitivity levels out as $t_G=2\kpg^2/I_s \gamma'^3\vt$, while the time the empty cavity sensitivity levels out is $t_E=2/(I_E\vt\kp)$.  The intersection time for when sensitivity with gain becomes better than that with an empty cavity is given by setting the empty case long time limit to the gain case intermediate time and is
\begin{eqnarray}
\frac{8\kpg^2}{c^2 I_s \gamma' t_c}&=&\frac{\kpe^2}{4c^2}\vt\\
t_c&=&\frac{2^5\kpg^2}{\kpe^2 I_s \gamma'\vt}.
\end{eqnarray}

We get the proper constraint by setting $\tx=\chi t_c<t_G$.  Note that this does not depend on $I_E$ because we assume the empty cavity performance hits the limit due to $\vt$ earlier than the crossover time, i.e. $t_E<t_c$.  We can rearrange $\tx=\chi t_c$ as
\begin{eqnarray}
\gamma'&=&\frac{2^5\kpg^2\chi}{\kpe^2 I_s \tx \vt}.
\end{eqnarray}
In other words, for a fixed $\vt$, for $\tx$ to be reasonably small and $\chi$ reasonably large, we need $\gamma'$ to be reasonably large and away from threshold (but still less than $\kpg$). We also need $\chi t_c < t_G$ or
\begin{eqnarray}
\chi &<& \frac{t_G}{t_c}=\frac{16 \kpe^2}{\gamma'^2}.
\end{eqnarray}
Numerically, this constraint describes the performance in the large $\vt$ limit.  In \reffig{Vt0}, the two sensitivities cross each other ($\chi=1$) when the technical noise variance is equal to
\begin{eqnarray}
\vt&=&\frac{2^5 \kpg^2}{\kpe^2 I_s \tx \gamma'}
\end{eqnarray}
Below the critical value of $\vt$, the empty cavity wins, and above the critical value the nonlinear method is preferred with an optimal  photon number that decreases as $\vt$ increases.    Here the sensitivity improves by the factor in the inequality above until the system approaches threshold ($\gamma'\approx \kpg/\sqrt{I_s}$ or $a\approx 1$) at which point the analysis becomes invalid (and does not diverge when made valid).  We also note that for small enough $\vt$ the empty cavity case levels off due to shotnoise.

\subsection{Gain with drive}

In this section, we demonstrate how spontaneous emission prevents gain added to a cavity with a drive field from producing an effectively `infinite finesse' cavity.  We have previously seen in \reffig{Heart} that the responsivity with drive does not do better than the zero-drive case, but it can outperform the empty cavity case in the presence of noise.  Here we aim to replicate the analysis of the last section with the goal of maximizing the sensitivity over both drive and gain for a fixed amount of measurement time.  In this case one can fill the cavity with the drive field to optimize the internal intensity, instead of relying on gain alone.  This may have practical consequences when considering technical constraints.

We follow the same linearization treatment as before, starting with \refeqn{GainSDE} and $Q=2\kg$.  However, we now set $\gamma'=(\kp-\kg)/2>0$ (so loss is slightly greater than gain) and also assume $\gamma'\ll\kp\approx\kg$.  To ignore the saturation term entirely, we require the condition $\kg I/ I_s \ll \kp-\kg = 2\gamma'$.  We can then linearize with $E_1=\bE_1+b_1$, $I=E_1^2=\tE_0^2/\gamma'^2$, where $\tE_0^2=\tE_{0,1}^2$ and $\tE_{0,2}^2=0$, giving
\begin{eqnarray}
db_1(t)&=&-\gamma' b_1dt+\sqrt{2Q}dW_i(t)
\end{eqnarray}
The variance is $\langle \Delta b_1^2 \rangle = 2Q/2\gamma'=2\kg/\gamma'$. For the linearized treatment, we need $I\gg \langle \Delta b_1^2 \rangle$, or
\begin{eqnarray}
\tE_0^2&\gg&2\kg\gamma'
\end{eqnarray}
and to ignore the saturation term we need 
\begin{eqnarray}
\kg I/I_s &\ll& 2\gamma'\\
\tE_0^2 &\ll& 2 I_s \gamma'^3 /\kg 
\end{eqnarray}
Note that this is more stringent than $I=\tE_0^2/\gamma'^2<I_s$ because $\gamma'/\kg\ll 1$.  Putting these conditions together we have
\begin{eqnarray}
2\kg\gamma' \ll \tE_0^2 \ll 2I_s\gamma'^3/\kg \\
\gamma' \gg \kg /\sqrt{I_s}
\end{eqnarray}
Thus for $\gamma'$ we have
\begin{eqnarray}
\kg/\sqrt{I_s} \ll \gamma' \ll \kg 
\end{eqnarray}
which gives
\begin{eqnarray}
\sqrt{I_s}\ll I=\tE_0^2/\gamma'^2 \ll I_s
\end{eqnarray}

Using $I=\tE_0^2/\gamma'^2$ and $\gamma'=(\kc+c\as-\kg)/2$, the responsivity is
\begin{eqnarray} 
\bar{R}&=&\frac{dI/d\as}{I}=(d\gamma'/d\as)\frac{dI/d\gamma'}{I}=\frac{-c}{\gamma'}
\end{eqnarray} 
which since $\gamma'\ll\kg$, can again be better than the empty cavity by a large factor.  The variance is given by $\langle \Delta I^2 \rangle=4 I \langle b_1^2 \rangle = 4(\tE_0^2/\gamma'^2) 2\kg/\gamma'=8\kg \tE_0^2/\gamma'^3$.  

We now minimize the sensitivity
\begin{eqnarray}
(\das^2)_G&=&\frac{1}{\bar{R}^2}\left(\frac{\langle \Delta I^2\rangle}{I^2}\frac{1}{1+\gamma't/2}+\vt\right)\nonumber\\
&=&\frac{\gamma'^2}{c^2}\left(\frac{8\kg \gamma'}{\tE_0^2}\frac{1}{1+\gamma' t/2}+\vt\right)\nonumber\\
&=&\frac{\gamma'^2}{c^2}\left(\frac{4\kp^2}{I_s \gamma'^2}\frac{1}{1+\gamma' t/2}+\vt\right)
\end{eqnarray}
where in the last step we maximize $\tE_0^2$ at its upper bound and use $\kg\approx\kp$.  This is essentially the same as \refeqn{GainUncertainty2} with the same limits on $\gamma'$.  Thus, when optimizing over $\gamma'$ at a specific time,  we observe the same trends as in \reffig{Vt0}: for large $\vt$ there is improvement optimized with $\gamma'$ small, and as $\vt$ decreases, the optimal $\gamma'$ increases and it gets more difficult to beat the empty cavity performance.

\subsection{Saturable Loss}

\begin{figure}[t]
\capstart
\includegraphics[width=3.0in]{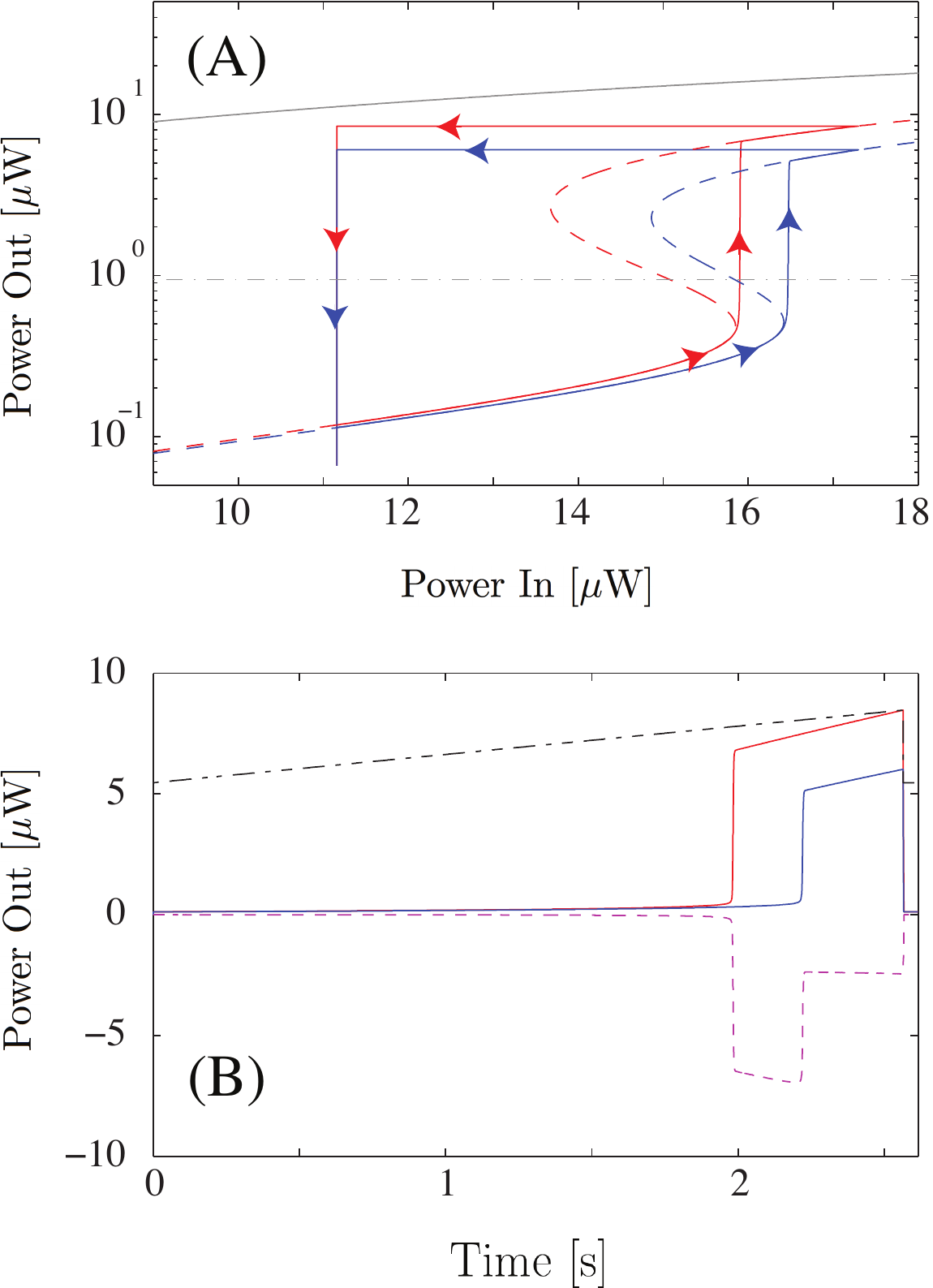}
\caption{The `sweep-up' modulation technique simulated for a $10^{-8} \ \cmi$ differential absorption signal and $\kl=12\kc$ (greater than the $8\kc$  needed for bistability).  (A) The output power (for the two cavities, red and blue) is shown versus the common input power, which is swept over the steady-state bistability curves (dashed lines).  The top gray line represents input equal to output.  (B) The two outputs (red and blue solid lines) as a function of time for the given input ramp (black dashed-dotted line). The differential output power (magenta dashed line) reveals the degree of differential absorption via the switching times.  Parameters are the same as in \reffig{Time}, but with an absorber having saturation intensity $I_{s, u}=1$ W/cm$^2$.} \label{Figure::AbsorptionGate}
\end{figure}

\begin{figure}[t]
\capstart
\includegraphics[width=3.0in]{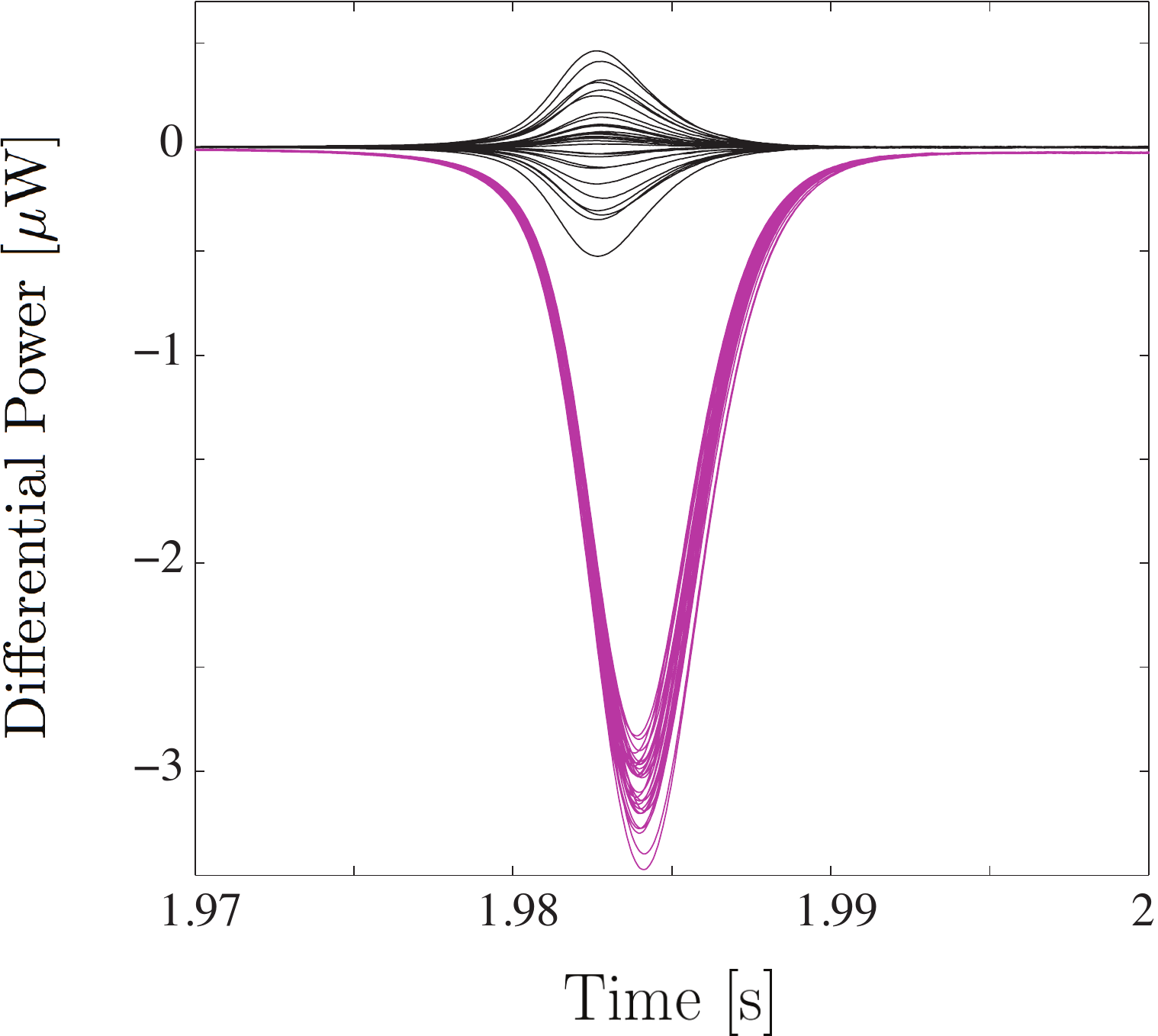}
\caption{As the differential absorption level is decreased the signal becomes obscured by spontaneous emission induced timing jitter.   For a signal of $10^{-10} \ \cmi$ (magenta) the differential blip is dominantly in the same direction (negative), yet when the signal is decreased to $10^{-12} \ \cmi$ (black) the order of hopping, hence the sign of the differential, becomes random, thus giving the noisefloor.  Parameters are the same as in \reffig{AbsorptionGate}.} \label{Figure::AbsorptionGateTime}
\end{figure}

In the gain-enhanced example above, the nonlinear kink near laser threshold as represented in \reffig{Heart} provides enhanced performance in the presence of technical noise for a steady-state measurement.  Within the same phase space, but operating with saturable loss rather than gain, increased responsivity is expected due to optical bistability, which occurs when the intracavity intensity is near saturation of an internal absorber, with the below-saturation absorber loss being much greater than the empty cavity loss \cite{Siegman1986, Drummond1980}.   

This system is analogous to a particle within a drive-dependent potential.  In the bistable region there exists a double-well potential, but if the cavity is driven too high or low, the potential is tilted such that only one local minimum exists.  In the quantum case, spontaneous emission induced tunneling between the wells occurs and the system equilibrates over long timescales with a single-valued mean for all drive fields.  The slope of the output versus input curve becomes large in the middle of the bistable region where the system spends an equal amount of time in each potential well.  Thus the variance is also large at this point corresponding to a large peak in responsivity \cite{Drummond1980}.  Note, however, that the tunneling rate becomes slow near this point, thus there is again a trade-off between fluctuation timescales and responsivity.

Implementing absorption spectroscopy with saturable loss requires temperature stabilizing a cavity with an internal saturable absorber such that the addition of the trace absorber kicks the intensity from one stable point to another in the bistable hysteresis curve.  In this section, we show, as in previous examples, that spontaneous emission limits the fundamental sensitivity from surpassing that achieved with CRDS, but once again, this method can make the system robust and the effective sensitivity superior to CRDS in the presence of technical noise.

We begin with the classical optical bistability equations and make the absorption intensity dependent through the relationship
\begin{eqnarray}
\kappa_L(I)&=&\frac{\kl}{1+I/I_s}.
\end{eqnarray}
From \refeqn{CavityIO} we have
\begin{eqnarray}
I&=&\frac{P_0}{\hbar\omega} \frac{\kc}{(\kc+\frac{\kl}{(1+I/I_s)}+c \as)^2}.
\end{eqnarray}
Plotting this equation gives the s-shaped optical bistability (hysteresis) curve.  The two internal intensities at which the slope is infinite are given by setting the derivative $dP_0/dI=0$ and solving which gives
\begin{eqnarray}
I_{\pm}&=&I_s \frac{\kl-2(\kc+c \as)\pm\sqrt{\kl(\kl-8(\kc+c\as))}}{2(\kc+c\as)}.
\end{eqnarray}
We define a unitless parameter $x\ll 1$ such that $\kl=8(\kc+c\as)(1+x^2)$, and $I_{\pm}\approx I_s(3\pm 4 x)$.  We see that the bistability occurs for $\kl>8\kc$ at values greater than three times the saturation intensity.  The resulting bistability curve exhibiting hysteresis is shown in \reffig{AbsorptionGate} for different levels of trace gas absorption.

Instead of repeating the steady-state treatment as in the examples above, we now consider a modulation based approach for several reasons.  First, the steady state Fokker-Planck solution cannot be integrated as above because of the intensity dependence of the spontaneous emission parameter $Q(I)=Q_0 I/(I+I_s)$.  Second, it is natural to compare to the CRDS standard which is also a modulation approach that is robust to several technical noise sources.  Third, we stress analogies to other modulated nonlinear metrology systems, such as fluxgate magnetometers, where the signal is sensed via a shift in a measured switching time.

We consider two potential modulation schemes.  First, we analyze a `sweep-up' scheme where the drive is ramped up and the difference in switching times measured.  Second, we compare ring-down schemes with and without saturable absorbers, and show that including the saturable absorbers allows the system to outperform empty cavity CRDS for certain technical noise models. In both cases, we simulate single stochastic trajectories using the SDE of \refeqn{GainSDE}, but with $\kg=-\kl<0$ and $Q(I)=Q_0 I/(I+I_s)$.

\subsubsection{Sweep-up with saturable absorbers}

We define the sweep-up modulation scheme as follows: we consider a measurement scheme with two cavities mounted in parallel on a monolithic block, and admit the trace absorber into one cavity but not the other. Alternatively, with a broadband saturable absorber but a narrow sample absorber one can use two frequency beams, one resonant and one far detuned from the sample. Both cavities or frequencies are driven with the same laser source dithered across the entire dual-valued part of the S-shaped response curve shown in \reffig{AbsorptionGate}.  As the drive field is swept to increasing values, the intracavity intensity will jump to the higher branch at different times due to the absorption differential from the trace gas.  

In this modulation scheme, the sweep rate is constrained to be faster than the tunneling time between stable points such that the intensity can be deterministically swept to the jump point without the internal intensity randomly switching states.  But the sweep rate is slow compared to the ring-down rate of the empty cavity $\kc$, or else the jump will be smooth.  Simulation results for this scheme are shown in \reffig{AbsorptionGate}, where the input intensity is swept over the bistability region at a moderate speed.  For the large differential absorption used in the plot,  the difference in switching times is represented clearly in the difference of output intensities.  The differential intensity for smaller signals is shown in \reffig{AbsorptionGateTime}.  If the differential absorption is large enough, the differential signal always has the same sign.   But as the absorption signal falls below a certain value, the order of the switching times become randomized due to spontaneous emission induced tunneling.  This scheme is analogous to the fluxgate magnetometer design, wherein a small field is sensed via the differential switching times of two oppositely driven solenoids containing ferromagnets \cite{Primdahl1979}.

\subsubsection{Ring-down with saturable absorbers}

Another modulation technique involves loading the cavities with an intensity much greater than the saturation intensity and monitoring the differential intensity decay or ring-down curve as in CRDS.  As opposed to CRDS, the evolution with saturable absorption displays two different timescales corresponding to the loss rates both above and below saturation.  The system starts decaying at a rate $\kc$ as if the absorber were absent, but once the intensity hits the saturation level, the absorber turns on and the decay becomes much larger, $\kc+\kl$, as shown in \reffig{AbsorptionGateCRDS}A.  The transition time between these two regimes depends on the internal trace absorber, and this time can be estimated by recording when the intensity drops below a certain sub-saturation threshold as compared to the other cavity.

\begin{figure}[t]
\capstart
\includegraphics[width=3.0in]{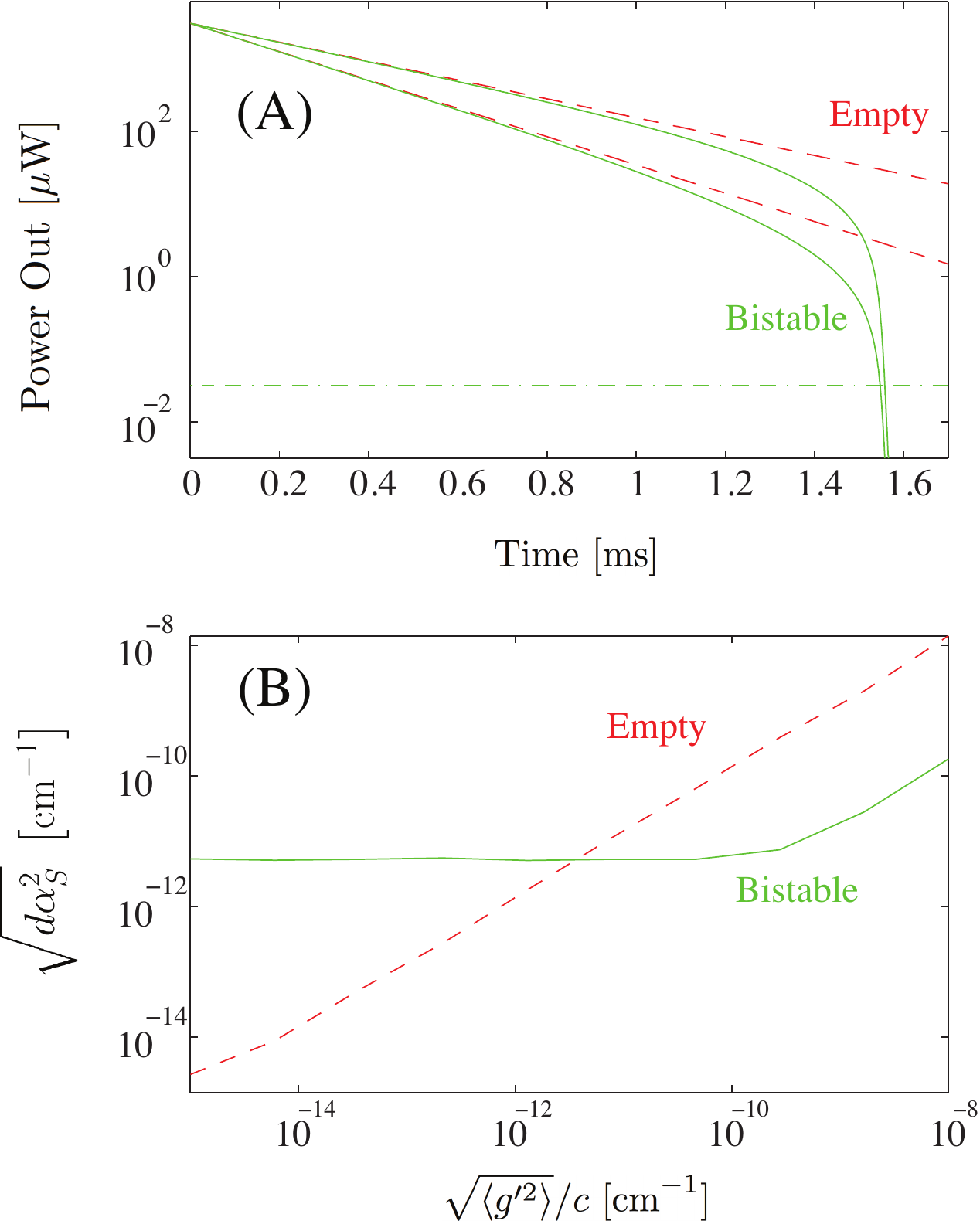}
\caption{(A) The dashed red curves show empty cavity ringdown as a function of time (in semilog scale) with `measurement noise' separation of $g'=0.5 \kc$ using the model described in the text.  The same measurement separation is shown for the saturable CRDS in the solid green curves.  The absorption signal is the same in each case and, in the saturable case, estimated from the times that the curves pass through a threshold intensity (green dashed-dotted curve at $I_s/10$).  (B) The performance from fitting regular CRDS (red dashed) and timing estimation with saturable CRDS (green solid).  Because the green curve timing is relatively immune to the multiplicative measurement noise, it performs better for large amounts of this noise where the CRDS estimation is limited by the indistinguishable measurement noise.  For low amounts of noise, the red curve hits the shotnoise limit and the green curve hits the limit from \refeqn{BadTimes}.  These approach each other as the saturation intensity approaches the mirror limit intensity.  We use the same cavity parameters as in \reffig{AbsorptionGateTime}, but with $\kl=100\kc$, $I_{s, u}=1$ W/cm$^2$, and initial intensity at the mirror limit of $I_{\textrm{init},u}=10$ kW/cm$^2$.} \label{Figure::AbsorptionGateCRDS}
\end{figure}

Again the timing performance will be limited by the spontaneous emission induced random tunneling.  We determine the absorption estimation error from the timing error as follows.  The intensity starts at $I_{\textrm{init}}$, and decreases past $I_s$ at time $t$ from which we can measure the absorption from $\alpha_S=-(\log(I_{\textrm{init}}/I_s)/ct)-\kc/c$.  When we are ramping up or down near $I_s$ at rate $\kc$ the variation $dI$ will trigger the jump process in a time window of width $dt=dI/I_s\kc$.  Thus the total sensitivity on a single shot will be roughly
\begin{eqnarray}
\sqrt{\das^2}&\approx&\log\left(\frac{I_{\textrm{init}}}{I_s}\right)\frac{dt}{ct^2}\approx \log\left(\frac{I_{\textrm{init}}}{I_s}\right)\frac{\kc}{c}\frac{dI}{I_s}\approx\frac{\kc}{c}\frac{1}{\sqrt{I_s}}\label{Equation::BadTimes}
\end{eqnarray}
where we have used $t\approx 1/\kc$ and $dI\approx\sqrt{I_s}$, $\log(I_{\textrm{init}}/I_s)\approx 1$.  This is roughly the same as the fundamental limit of \refeqn{EmptyUncertainty2}, but with $I_s$ instead of the $I_E$.  Because $I_s$ cannot surpass $I_E$ this limit is generally worse, but we see that saturable absorbers with high $I_s$ are preferable.  This same argument can be applied to the sweep-up case of the last section as well.

We next compare the robustness of this saturable ring-down method to empty cavity CRDS for a particular noise model.  We propose a certain multiplicative measurement noise model that takes the output light and multiplies it by a factor $\exp(g' t)$, where `gain drift' $g'$ in inverse time units has mean zero but finite variance and is constant within a single shot (i.e., $\langle g'^2 \rangle\neq 0$, $\langle g' \rangle =0$).  This model represents noise due to a time-dependent gain in the measurement process which could, for example, be due to actual gain fluctuations in the detection or alignment instabilities.  This is clearly seen when the exponential is expanded to give $1+g't$.  However, the exponential form of the model was chosen such that this noise process is fundamentally indistinguishable from the cavity decay in CRDS.  In other words, the noise takes $\kc\rightarrow\kc-g'$ and thus no fitting procedure can distinguish $\kc$ from $g'$ in a single-shot.  Thus, the estimation error is equal to $\langle g'^2 \rangle$ when this is greater than the shotnoise limit.  In contrast, the nonlinear ring-down method displays a kink which is not shifted substantially in time when this noise model is applied.  The threshold value from which the timing is taken should be below the switching point where the intensity becomes less than saturation $I < I_s$.  Past this point the loss curve is sharper vertically by a factor of $\kl/\kc\gg 1$, hence the timing is robust to such time-dependent gain errors by this same factor. The resulting robustness of the nonlinear estimator is compared to the non-robust CRDS in \reffig{AbsorptionGateCRDS}.  The timing estimator is robust enough to outperform CRDS for large values of $\langle g'^2 \rangle$ but the error does begin to rise at extremely large values of this noise.

This multiplicative measurement noise model is generally representative of any time-dependent measurement gain during the time of the ring-down.  One could propose another noise model that moves the kink location in the nonlinear case as well, but such a model is practically less likely for slow noise because it relies on having at least a quadratic noise signal in time, requiring a higher order expansion of the noise source.  We stress that this improved robustness is only for a particular technical noise model on the measurement side.  The CRDS method remains relatively immune to initial intensity fluctuations, whereas the saturable-absorber method does not.

\section{Conclusions}

In this paper, we have analyzed nonlinear methods for high-sensitivity, intracavity absorption spectroscopy.  Although inherent quantum noise mechanisms, such as spontaneous emission and shotnoise, prevent the fundamental limits of nonlinear approaches from surpassing those of linear methods (e.g., empty cavity ringdown spectroscopy), we have shown that they present an advantageous robustness to certain kinds of technical noise.  This enhancement is due to nonlinearities of the system response which can be a function of many experimentally convenient parameters, such as drive, gain, or time.

Implementing these nonlinear techniques involves additional experimental costs and challenges, which may ultimately introduce larger technical noise sources.  The gain and saturable-absorber systems will have some degree of technical fluctuations in the medium (e.g., due to thermal atom number fluctuations) and also will likely reduce the cavity finesse due to insertion loss.  However, specific implementations may warrant the effort when considering commercial, field-deployable devices where environmental noise sources may be significant.  

\section*{Acknowledgements}

We thank Hideo Mabuchi and Michael Armen for comments and acknowledge A. Kushner and J. Kushner.

\bibliography{Bib}

\end{document}